\documentclass[conference]{IEEEtran}
\IEEEoverridecommandlockouts
\usepackage{cite}
\usepackage{subcaption}
\usepackage{url}

\newcommand{\squeezeup}{\vspace{-3mm}}

\usepackage{amsmath,amssymb,amsfonts}
\usepackage{algorithmic}
\usepackage{graphicx}
\usepackage{textcomp}
\usepackage{xcolor}
\usepackage{float}    

\usepackage{glossaries} 
\newacronym{adc}{ADC}{analogue to digital converter}
\newacronym{asm}{ASM}{attached synchronous marker}
\newacronym{apd}{APD}{Avalanche Photodiode}
\newacronym{adcs}{ADCS}{Attitude Determination and Control System}
\newacronym{agn}{AGN}{Active Galactic Nuclei}
\newacronym{ascot}{ASCOT}{The Advanced Scintillator Compton Telescope}
\newacronym{asic}{ASIC}{application specific integrated circuit}
\newacronym{ata}{ATA}{Admission Temporaire/Temporary Admission}
\newacronym{batse}{BATSE}{Burst And Transient Source Experiment}
\newacronym{bga}{BGA}{Ball Grid Array}
\newacronym{cebr}{CeBr$_3$}{cerium bromide}
\newacronym{cgro}{CGRO}{Compton Gamma Ray Observatory}
\newacronym{ci}{CI}{Current Integrator}
\newacronym{cmis}{CMIS}{Current Mode Input Stage}
\newacronym{cob}{CoB}{Chip on Board}
\newacronym{cots}{COTS}{Commercial Off-The-Shelf}
\newacronym{crf}{CRF}{Coordinate Reference Frame}
\newacronym{cpld}{CPLD}{Complex Programmable Logic Device}
\newacronym{ccs}{CCS}{Code Composer Studio}
\newacronym{csbf}{CSBF}{Columbia Scientific Balloon Facility}
\newacronym{cad}{CAD}{Computer Aided Design}
\newacronym{csf}{CSF}{CubeSat Support Facility}
\newacronym{cpu}{CPU}{central processing unit}
\newacronym{csi}{CsI}{Caesium Iodide}
\newacronym{crc}{CRC}{cyclical redundancy check}
\newacronym{dac}{DAC}{Digital to Analogue Converter}
\newacronym{dil}{DIL}{Dual In-Line}
\newacronym{dm}{DM}{Demonstration Model}
\newacronym{dma}{DMA}{direct memory access}
\newacronym{ddf}{DDF}{Design Definition File}
\newacronym{dof}{DOF}{Degree of Freedom}
\newacronym{em}{EM}{electromagnetic}
\newacronym{egse}{EGSE}{Electrical Ground Support Equipment}
\newacronym{eqm}{EQM}{Engineering Qualification Model}
\newacronym{esa}{ESA}{European Space Agency}
\newacronym{eirsat}{EIRSAT-1}{Educational Irish Research Satellite-1}
\newacronym{eusci}{eUSCI}{enhanced universal serial communication interface}
\newacronym{fwhm}{FWHM}{Full Width Half Maximum}
\newacronym{fm}{FM}{Flight Model}
\newacronym{fe}{FE}{Finite Element}
\newacronym{fys}{FYS}{Fly Your Satellite!}
\newacronym{ftdi}{FTDI}{Future Technology Devices International}
\newacronym{fram}{FRAM}{ferroelectric random access memory}
\newacronym{gecam}{GECAM}{Gravitational Wave High-Energy Electromagnetic Counterpart All-Sky Monitor}
\newacronym{gbm}{GBM}{Gamma-Ray Burst Monitor}
\newacronym{gifts}{GIFTS}{Gamma-Ray Investigation of the Full Transient Sky}
\newacronym{gmod}{GMOD}{Gamma-Ray Module}
\newacronym{gmodem}{GMoDem}{Gamma-Ray Module Demonstration}
\newacronym{gnss}{GNSS}{Global Navigation Satellite System}
\newacronym{gpio}{GPIO}{general purpose input output}
\newacronym{gps}{GPS}{Global Positioning System}
\newacronym{grb}{GRB}{gamma-ray burst}
\newacronym{grd}{GRD}{Gamma-Ray Detector}
\newacronym{grid}{GRID}{Gamma-Ray Integrated Detectors}
\newacronym{grips}{GRIPS}{Gamma-Ray Imaging, Polarimetry and Spectroscopy}
\newacronym{gsfc}{GSFC}{Goddard Space Flight Center}
\newacronym{gui}{GUI}{Graphical User Interface}
\newacronym{grm}{GRM}{Gamma-Ray Monitor}
\newacronym{gw}{GW}{gravitational wave}
\newacronym{gse}{GSE}{Ground Support Equipment}
\newacronym{hdl}{HDL}{Hardware Description Language}
\newacronym{hpge}{HPGe}{High Purity Germanium}
\newacronym{htv}{HTV}{H-II Transfer Vehicle}
\newacronym{integral}{INTEGRAL}{International Gamma-Ray Astrophysics Laboratory}
\newacronym{iut}{IUT}{Item Under Test}
\newacronym{iss}{ISS}{International Space Station}
\newacronym{ide}{IDE}{integrated development environment}
\newacronym{lpm}{LPM}{low power mode}
\newacronym{labr}{LaBr$_3$}{Lanthanum Bromide}
\newacronym{led}{LED}{Light Emitting Diode}
\newacronym{leo}{LEO}{low earth orbit}
\newacronym{ligo}{LIGO}{Laser Interferometer Gravitational-Wave Observatory}
\newacronym{lvttl}{LVTTL}{Low Voltage Transistor Transistor Logic}
\newacronym{liso2}{LiSO$_2$}{Lithium Sulfur Dioxide}
\newacronym{i2c}{I$^{2}$C}{Inter-Integrated Circuit}
\newacronym{ic}{IC}{Integrated Circuit}
\newacronym{ideas}{IDEAS}{Integrated Detector Electronics AS}
\newacronym{ipa}{IPA}{Isopropyl Alcohol}
\newacronym{ir}{IR}{Infrared}
\newacronym{insight}{InSight}{Interior Exploration using Seismic Investigations, Geodesy and Heat Transport}
\newacronym{jtag}{JTAG}{Joint Test Action Group}
\newacronym{mos}{MoS}{Margin of Safety}
\newacronym{moonbeam}{MoonBEAM}{Moon Burst Energetics All-sky Monitor}
\newacronym{mb}{MB}{motherboard}
\newacronym{mppc}{MPPC}{Multi-Pixel Photon Counter}
\newacronym{marco}{MarCO}{Mars Cube One}
\newacronym{megalib}{MEGAlib}{The Medium-Energy Gamma-Ray Astronomy Library}
\newacronym{nai}{NaI}{Sodium Iodide}
\newacronym{ncr}{NCR}{Non-Conformance Report}
\newacronym{nasa}{NASA}{National Aeronautics and Space Administration}
\newacronym{obc}{OBC}{on-board computer}
\newacronym{opamp}{op-amp}{Operational Amplifier}
\newacronym{pcb}{PCB}{printed circuit board}
\newacronym{ptr}{PTR}{Post Test Review}
\newacronym{pmt}{PMT}{photomultiplier tube}
\newacronym{pps}{PPS}{Pulse Per-Second}
\newacronym{psu}{PSU}{power supply unit}
\newacronym{tgf}{TGF}{Terrestrial Gamma-Ray Flash}
\newacronym{tte}{TTE}{time-tagged event}
\newacronym{tsv}{TSV}{Through Silicon Via}
\newacronym{tvac}{TVAC}{Thermal-Vacuum Chamber}
\newacronym{ucd}{UCD}{University College Dublin}
\newacronym{unh}{UNH}{University of New Hampshire, Durham, USA}
\newacronym{ui}{UI}{User Interface}
\newacronym{usb}{USB}{Universal Serial Bus}
\newacronym{uart}{UART}{Universal Asynchronous Receiver/Transmitter}
\newacronym{rbf}{RBF}{Remove Before Flight}
\newacronym{rga}{RGA}{Residual Gas Analyzer}
\newacronym{rft}{RFT}{Reduced Functional Test}
\newacronym{risc}{RISC}{Reduced Instruction Set Computer}
\newacronym{s3}{S$^{3}$}{Small Space Simulator}
\newacronym{sar}{SAR}{successive approximation register}
\newacronym{sbc}{SBC}{Single-board Computer}
\newacronym{sbw}{SBW}{Spy-Bi-Wire}
\newacronym{sbd}{SBD}{Short Burst Data}
\newacronym{saa}{SAA}{South Atlantic Anomaly}
\newacronym{siphra}{SIPHRA}{Silicon Photomultiplier Readout ASIC}
\newacronym{sipm}{SiPM}{silicon photomultiplier}
\newacronym{spad}{SPAD}{Single Photon Avalanche Diode}
\newacronym{spi}{SPI}{Serial Peripheral Interface}
\newacronym{sip}{SIP}{Shaker Interface Plate}
\newacronym{sap}{SAP}{Subsystem Adaptor Plate}
\newacronym{ser}{SER}{soft error rate}
\newacronym{sram}{SRAM}{static random access memory}
\newacronym{tstp}{TSTP}{Test Specification - Test Procedure}
\newacronym{tspe}{TSPE}{Test Specification}
\newacronym{tpro}{TPRO}{Test Procedure}
\newacronym{trpt}{TRPT}{Test Report}
\newacronym{trp}{TRP}{Temperature Reference Point}
\newacronym{trr}{TRR}{Test Readiness Review}
\newacronym{tqcm}{TQCM}{Thermoelectric Quartz Crystal Microbalance}
\newacronym{ti}{TI}{Texas Instruments}
\newacronym{vhdl}{VHDL}{Very High Speed Integrated Circuit Hardware Description Language}
\newacronym{ptfe}{PTFE}{PolyTetraFluoroEthylene}
\newacronym{xsvf}{XSVF}{Xilinx Serial Vector Format}

\def\BibTeX{{\rm B\kern-.05em{\sc i\kern-.025em b}\kern-.08em
    T\kern-.1667em\lower.7ex\hbox{E}\kern-.125emX}}
    
\begin{document}
    
\makeatletter
\def\footnoterule{\kern-3\p@
  \hrule \@width 2in \kern 2.6\p@} 
\makeatother

\makeatletter
    \newcommand{\linebreakand}{%
      \end{@IEEEauthorhalign}
      \hfill\mbox{}\par
      \mbox{}\hfill\begin{@IEEEauthorhalign}
    }
    \makeatother

\title{Embedded Firmware Development for a Novel CubeSat Gamma-Ray Detector\\}

\author{\IEEEauthorblockN{1\textsuperscript{st} Joseph Mangan\textsuperscript{*}}
\IEEEauthorblockA{\textit{School of Physics} \\
\textit{University College Dublin,}\\ 
Dublin, Ireland \\
joseph.mangan\\
@ucdconnect.ie}
\thanks{*This work is supported by Science Foundation Ireland (SFI) under grant number 17/CDA/4723 and supported by The European Space Agency's Science Programme under contract 4000104771/11/NL/CBi.
} 

\and
\IEEEauthorblockN{2\textsuperscript{nd} David Murphy}
\IEEEauthorblockA{\textit{School of Physics} \\
\textit{University College Dublin}\\
Dublin, Ireland \\
david.murphy@ucd.ie\\
}
\and

\IEEEauthorblockN{3\textsuperscript{rd} Rachel Dunwoody}
\IEEEauthorblockA{\textit{School of Physics} \\
\textit{University College Dublin}\\
Dublin, Ireland \\
rachel.dunwoody\\
@ucdconnect.ie}

\and
\IEEEauthorblockN{4\textsuperscript{th} Maeve Doyle}
\IEEEauthorblockA{\textit{School of Physics} \\
\textit{University College Dublin}\\
Dublin, Ireland \\
maeve.doyle.1\\
@ucdconnect.ie}
\and
\linebreakand

\IEEEauthorblockN{5\textsuperscript{th} Alexey Ulyanov}
\IEEEauthorblockA{\textit{School of Physics} \\
\textit{University College Dublin}\\
Dublin, Ireland \\
alexey.uliyanov@ucd.ie}
\and
\IEEEauthorblockN{6\textsuperscript{th} Lorraine Hanlon}
\IEEEauthorblockA{\textit{School of Physics} \\
\textit{University College Dublin}\\
Dublin, Ireland \\
lorraine.hanlon@ucd.ie}
\and
\IEEEauthorblockN{7\textsuperscript{th} Brian Shortt}
\IEEEauthorblockA{\textit{European Space Agency} \\
\textit{ESTEC}, \\
\textit{Netherlands} \\
brian.shortt@esa.int}
\and
\IEEEauthorblockN{8\textsuperscript{th} Sheila McBreen\textsuperscript{**}}
\IEEEauthorblockA{\textit{School of Physics} \\
\textit{University College Dublin}\\
Dublin, Ireland \\
sheila.mcbreen@ucd.ie}
\thanks{\textit{**The full EIRSAT-1 team also includes Masoud Emam, Jessica Erkal, Joe Flanagan, Gianluca Fontanesi, Andrew Gloster, Conor O'Toole, 
Favour Okosun, Rakhi RajagopalanNair, Jack Reilly, Lána Salmon, Daire Sherwin, Joseph Thompson, Sarah Walsh, Daithí de Faoite, Mike Hibbett, Umair Javaid, Fergal Marshall, David McKeown, William O'Connor, Kenneth Stanton, Ronan Wall
}}
}
\maketitle

\begin{abstract}
The Gamma-ray Module (GMOD) is an experiment designed for the detection of gamma-ray bursts in low Earth orbit as the principal scientific payload on a 2-U CubeSat, EIRSAT-1. GMOD comprises a cerium bromide scintillator coupled to silicon photomultipliers which are processed and digitised by a bespoke ASIC. Custom firmware on the GMOD motherboard has been designed, implemented and tested for the MSP430 microprocessor which manages the experiment including readout, storage and configuration of the system. The firmware has been verified in a series of experiments testing the response over a realistic range of input detector trigger frequencies from 50\,Hz to 1\,kHz for the primary time tagged event (TTE) data. The power consumption and ability of the firmware to successfully receive and transmit the packets to the on-board computer was investigated. The experiment demonstrated less than 1\,\% loss of packets up to 1\,kHz for the standard transfer mode with the power not exceeding 31\,mW. The transfer performance and power consumption demonstrated are within the required range of this  CubeSat instrument. 

\end{abstract}

\vspace{-0.1cm}

\section{Introduction}

The \gls{gmod}\cite{Murphy_2021} is a novel instrument developed to study high energy astrophysical transient events called \glspl{grb}\cite{Vedrenne_Atteia_2010} in the era of multi-messenger astronomy. \gls{gmod} will be flown in \gls{leo} on-board the \gls{eirsat}, a 2U CubeSat developed as part of the \gls{esa} \gls{fys} programme\cite{Murphy_2018}. The sustained miniaturisation of technology has been the driving force behind the development of CubeSats for science applications\cite{OnthevergeofanastronomyCubeSatrevolution}; these are small satellites whose mass and dimensions are defined in units of ``U'' where 1U corresponds to a CubeSat of 10cm$\times$10cm$\times$10cm in volume and approximately 1.3kg in mass\cite{CDS_2014}. \gls{gmod} is a $<$1U scintillation based gamma-ray detecting instrument which features modern miniaturised detector technology in a robust and low power, CubeSat compatible form factor. The development of the \gls{gmod} detector builds on previous work including a number of technology investigations \cite{Ulyanov_2016,Ulyanov_2017a}, a five hour balloon flight at 37km float altitude \cite{Murphy_2021_balloon}, a 101.4MeV proton irradiation test \cite{Ulyanov_2020} and most recently, the engineering qualification model environmental test campaign, including vibration and thermal-vacuum testing \cite{Mangan_2021}. \gls{gmod} is expected to detect $\sim$11--14 \glspl{grb} over a one year mission at a significance of 10\,$\sigma$, within the standard 50\,keV to 300\,keV energy range \cite{Murphy_2021_a}.

When complete, \gls{gmod} will be used as the basis for a scaled up configuration of CubeSat compatible gamma-ray detectors allowing localisation capabilities as part of the \gls{gifts} project. \gls{gmod} and other CubeSat instruments can also be used to augment the observing capabilities of the current fleet of \gls*{grb} detecting missions. These include large missions such as the Fermi Space Telescope\cite{Meegan_2009}, INTEGRAL\cite{Winkler_2003}, the Neil Gehrels Swift Telescope and planned CubeSat missions and small satellites such as BurstCube\cite{Racusin_2016} and \acrshort{gecam}\cite{ZHANG20198}. 

In 2017, a breakthrough discovery was made when GRB170817A, a coincident electromagnetic counterpart to GW170817, a \gls{gw} event was detected \cite{Abbott_2017,Goldstein_2017}. This coincident detection marked the beginning of a new era of multi-messenger astronomy and highlighted the importance of gamma-ray missions in the detection of coincident \gls*{gw} and \gls{grb} events, an effort which \gls*{gmod} will contribute to as a member of the wider \gls*{grb} community.

In addition to the science goals of detecting \glspl{grb}, \gls{gmod} will further the technology readiness of advanced detector technologies; including that of \glspl{sipm} which are used to provide a compact, low power, and rugged solid state alternative to the larger, high-voltage and fragile \glspl{pmt}, as used on earlier space borne gamma-ray detecting instruments. The \gls{gmod} detector will be functionally supported by a custom built readout and support system consisting of a \gls{mb} which will provide a number of operation critical duties. This in-house developed system will accomplish this using a C/C++ firmware running on a Texas Instruments MSP430FR5994 mixed signal processor, a 16-bit, low power, 16\,MHz microcontroller. We present an overview of the development of this firmware, the context of its requirements for satisfactory operation and an initial assessment of its performance to date with planned improvements and future work.

\begin{figure*}[t!]
\centerline{\includegraphics[width=0.77\linewidth]{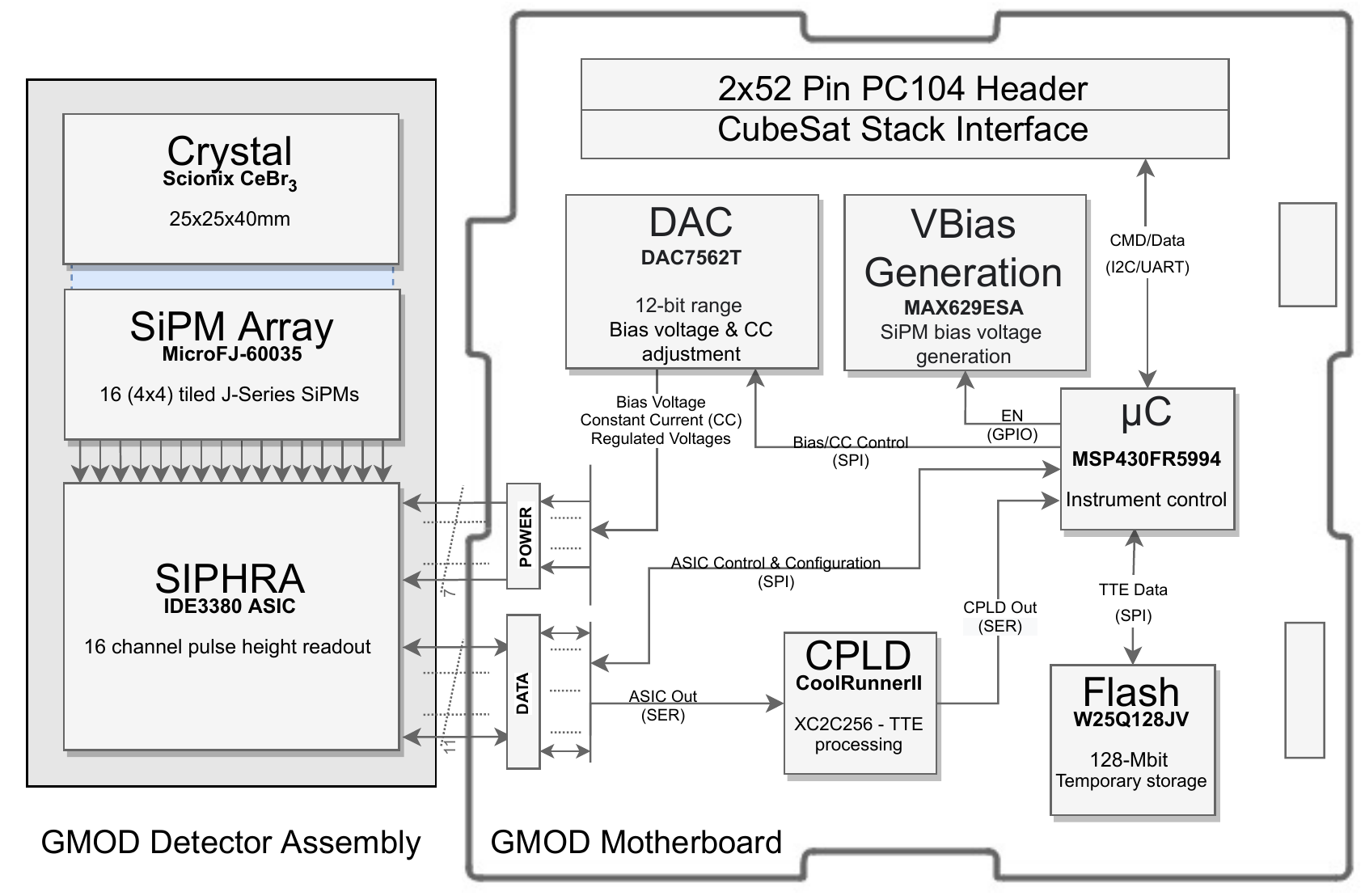}}
\caption{The high level operational diagram of GMOD, including the detector assembly, GMOD motherboard and principal components. For the sake of clarity only the most relevant details are encompassed in this diagram. A complete overview of the instrument hardware is presented in \cite{Murphy_2021}.}
\squeezeup
\label{fig:GMOD_flow}
\end{figure*}

\section{The Gamma-ray Module}

\gls{gmod} comprises a 25mm$\times$25mm$\times$40mm Scionix \gls{cebr} scintillator which is optically coupled to a custom built tiled array of 16 (4$\times$4) OnSemiconductor MicroFJ-60035-TSV \glspl{sipm} \cite{Murphy_2021}\cite{Murphy_2018}\cite{Murphy_2021_a}. Front-end readout for the \gls{sipm} array is provided by the 16 channel IDE3380 SIPHRA \gls{asic}, produced by IDEAS (Norway)\cite{SIPHRA,Ulyanov_2017b}. The \gls{asic} was developed to be a readout solution for pulse height spectroscopy in instruments using \glspl{sipm}, \glspl{pmt} or multi-pixel photon counters and is implemented as a bare silicon die wire-bonded to a carrier board \gls{pcb}. This detector assembly is sealed within a light-tight custom machined aluminium enclosure and bolted directly to the \gls{gmod} \gls{mb}. The \gls{gmod} \gls*{mb} is a bespoke readout and support system for the \gls{gmod} detector assembly, providing a number of operation critical duties, including the generation of an adjustable \gls{sipm} bias voltage, a constant current source, system configuration, readout and the temporary on-board flash storage of individual gamma-ray \gls{tte} packets. This was developed over a number of demonstration model iterations to be both a tailored solution for the instruments readout and control, as well as to serve as a structural mounting point for the detector assembly hardware. A high level operational layout of the \gls{gmod} instrument is presented in Figure\,\ref{fig:GMOD_flow}, while a component list is detailed in Table\,\ref{tab:GMOD_specs_table}. 

Readout of the detector occurs as follows: a gamma-ray is either partially or fully absorbed by the \gls*{cebr} scintillator, which in turn produces a number of photons proportional to the energy of the incident gamma-ray within a wavelength matching the peak photo sensitivity of the J-Series \gls{sipm} array. Each \gls*{sipm} is made up of an array of many thousands of microcells, avalanche photodiodes in Geiger-mode operation with a series quench resistor; photons absorbed by these microcells trigger an avalanche flow releasing a fixed amount of charge which is proportional to the energy of the incident gamma-ray photon. SIPHRA integrates and shapes this current signal, which when surpassing a fixed trigger threshold, triggers digitisation of the pulse height using a 12-bit \gls{sar} \gls{adc}. The \gls{asic} then outputs the event data as a 20-bit serial stream at 1M\,Baud for each active \gls{sipm} channel, 12-bits for the \gls{adc} measurement and the remaining bits allocated to channel trigger information. This is an unusual form of serial transmission which cannot be readily handled by the MSP430 \gls{eusci} module and must first be reduced for efficient bit packing. This is performed by a Xilinx XC2C256 \gls{cpld} which directly receives the SIPHRA trigger data and produces two types of \gls{tte}: 16 channel, containing measurements for each active \gls{sipm} channel and summed channel, containing the summed signal measurements from the \gls{sipm} array. The \gls{cpld} also applies a 4 byte fine-timestamp with a 1\,$\mu$s time base and an \gls{asm} as identification bytes for distinct recognition of the \gls{tte} type (as either summed channel or 16 channel data). It is at this point that both summed and 16 channel \glspl{tte}, consisting of the pulse height spectroscopic measurements from the detector, are transmitted to the MSP430FR5994 microcontroller over 1\,MBaud serial for processing and temporary storage, before eventual transmission to the \gls{eirsat} \gls{obc}\cite{Doyle_2020}. 

According to simulations\cite{Murphy_2021_a} performed using the \gls{megalib}\cite{Zoglauer_2006}, \gls{gmod} is expected to have a background gamma-ray trigger rate of $\sim$50\,Hz within the 50 - 300\,keV energy range, and (possibly closer to $\sim$100\,Hz in a wider energy band). \gls{gmod} is expected to detect between 11 and 14 \glspl{grb} over its one year mission at a significance of 10\,$\sigma$, depending on spacecraft pointing\cite{Murphy_2021_a}. This corresponds to a SIPHRA trigger rate of around $\sim$120\,Hz over a single 1024\,ms binned light curve record. It is not expected that \gls{gmod} will see \glspl{grb} which generate over 500\,counts per second and the vast majority of \gls{grb} triggers will be less than 10\,$\sigma$. In terms of the extreme, SIPHRA's trigger rate in \gls{gmod} can just about exceed 1000\,Hz, after which it is capped by the time it takes to output all active \gls{sipm} channel readout. An event triggering SIPHRA at such high rates for any long duration is not expected, however high rates are expected while transiting the \gls{saa} (due to the concentration of high energy protons) or if passing through high latitude regions (due to the concentration of high energy electrons)

\newcommand\cola{2cm}
\newcommand\colb{3.09cm}
\begin{table}
	\caption{A component list of the GMOD instrument, including a high level overview of the detector and the GMOD motherboard hardware and a list of the most relevant MSP430FR5994 internal peripheral modules.} \label{tab:GMOD_specs_table}
	\begin{center}
	\begin{tabular}{p{0.005cm} p{0.75cm} l l l} 
	\hline \hline \\ [-1.5ex]
    \multicolumn{5}{l}{Detector}\\
             & Crystal &  \multicolumn{3}{l}{Scionix \gls{cebr} 25$\times$25$\times$40mm}\\
             & SiPM   &  \multicolumn{3}{l}{OnSemiconductor MicroFJ60035-TSV}\\
             & Array        &  \multicolumn{3}{l}{Custom 4x4 (16$\times$) tiled SiPM array}\\
             & Front-end    &  \multicolumn{3}{l}{SIPHRA IDE3380 ASIC (Carrier Board)}\\
             \hline\\ [-1.5ex]
    \multicolumn{5}{l}{Motherboard}\\
             & $\mu$C       &  \multicolumn{3}{l}{16-bit, Texas Instruments MSP430FR5994}\\
             & CPLD   &  \multicolumn{3}{l}{Xilinx CoolRunnerII XC2C256}\\
             & Storage      &  \multicolumn{3}{l}{128-Mbit (16\,MB) temporary flash storage}\\
             & DAC          &  \multicolumn{3}{l}{12-bit DAC7562T (bias/CC supply adjustment)} \\
             \hline\\ [-1.5ex]
    \multicolumn{5}{l}{MSP430FR5994 Peripherals}\\
    \\ [-1.5ex]
             & CPU          &  \multicolumn{1}{p{\cola}|}{\raggedright{16\,MHz, 16-bit RISC architecture}}          & Clk Sys& \multicolumn{1}{p{\colb}}{\raggedright{3$\times$ independent clocks (MCLK, SMCLK, ACLK)}}\\
             & FRAM         &  \multicolumn{1}{p{\cola}|}{\raggedright{256\,kB storage}}                             & ADC & \multicolumn{1}{p{\colb}}{\raggedright{12-bit SAR}}\\
             & SRAM         &  \multicolumn{1}{p{\cola}|}{\raggedright{8\,kB storage}}                               & GPIO& \multicolumn{1}{p{\colb}}{\raggedright{68 multiplexed I/O}}\\
             & DMA          &  \multicolumn{1}{p{\cola}|}{\raggedright{6 channels}}                                 & CRC & \multicolumn{1}{p{\colb}}{\raggedright{32-bit $\&$ 16-bit CRC}}\\
             & Timer        &  \multicolumn{1}{p{\cola}|}{\raggedright{6$\times$ 16-bit timers with 7$\times$ CCRs}}& eUSCI& \multicolumn{1}{p{\colb}}{\raggedright{eUSCI\_A (4$\times$SPI/UART), eUSCI\_B (4$\times$SPI/I2C)}}\\
             \\ [-1.5ex]
	\hline \hline
	\end{tabular} 
	\squeezeup
	\squeezeup
	\end{center}
\end{table}

\section{The MSP430FR5994}
At the core of this system is the \gls{ti} MSP430FR5994 \cite{SLAU367P_2012} \cite{SLASE54D_2016}, a 16\,MHz mixed signal microcontroller based on a reduced instruction set computer architecture. This device was selected for numerous reasons: its low-power optimisation, 256\,kB of high speed \gls{fram}, a variety of integrated peripheral modules, past family space heritage\cite{Schoolcraft_2017} and ease of accessibility. The MSP430 can be placed into an number of operational states called ``low power modes'', effectively a set of predefined states of activity which selectively enable or disable the \gls{cpu}, peripheral modules and their relevant clock sources. The current consumption of any digital device is in general related to the rate at which it is clocked; the MSP430 clock system can produce three independently configurable clock signals for the \gls{cpu} and peripherals. By varying the \gls{lpm} setting certain clock sources and peripheral resources may be placed into a low power state while others remain active. This allows the user to develop a firmware which can optimise the current consumption of the device depending on the stage of operation. The combined use of \gls{lpm} states and internal \gls{fram} storage helps to reduce the power requirements of the MSP430. \gls{fram} is equivalent to flash memory in terms of its interface and non-volatility, but superior in terms of read/write access (close to 8\,MBps compared to flash at 14\,kBps), read/write endurance (10$^{15}$ write/erase cycles compared to 10$^{5}$ for flash)\cite{SLAA498B_2011}, and ultra-low-power requirements (requiring just 1.5V compared to 10-14V for the flash charge pump). It has even been shown to have some level of radiation resilience\cite{Guertin_2015} and a reduced \gls{ser}\cite{SLAA526A_2014} as compared to \gls{sram}, 8\,kB of which is available on the MSP430FR5994\footnote{This is partitioned in the default linker file provided by \gls{ti} in \gls{ccs}, into 4\,kB blocks, half allocated to common use by the \gls{cpu} and stack, while the other reserved for the Low-Energy Accelerator module. As this is unused on \gls*{gmod}, this has been merged into a single 8\,kB block.}. The device has a large selection of internal peripheral modules: six 16-bit timers with multiple capture compare registers, two \gls{eusci} modules (eUSCI\_A supporting up to 4 channels independently configurable as either UART/SPI and eUSCI\_B supporting up to 4 channels independently configurable as either I2C/SPI), a 6 channel \gls{dma} controller, a 12-bit \gls{sar} \gls{adc} with multiplexed inputs and 68 multiplexed general purpose input/outputs.  
These combined make the MSP430 a versatile option for the multiple requirements of \gls{gmod} and the general use of it in a variety of space applications. For example, the 430 platform of the MSP family has seen frequent use within CubeSat Kits\cite{cubesat_kit_2021, gauss_2021}, small satellites\cite{Ubbels_2005}\cite{klesh_2013} and CubeSat subsystems\cite{Aslan_2014}. It was most notably used on the \gls{marco}\cite{Schoolcraft_2017} CubeSats, companions to the \gls{nasa} \gls{insight} lander during its 2018 Martian arrival. Finally, a large resource of example code\footnote{\url{https://www.ti.com/product/MSP430FR5994}} has been made available by \gls{ti} for the MSP family of devices, including the MSP430FR5994, which hosts both register level and driver level examples. A wealth of documentation and application notes are also available with detailed information on all aspects of the register configuration settings for device operation. This makes writing and testing firmware on the MSP430 an approachable task for CubeSat teams, many of which are composed of university level students, as is the case for \gls{eirsat}.

A firmware specifically designed to take advantage of all of these qualities can provide high performance with a low power optimised operation. A preliminary C/C++ firmware has been developed for \gls{gmod} using the Eclipse based, \gls{ti} \gls{ccs}\cite{css_2021} \gls{ide}. To date, the firmware has been verified in a series of experiments testing its response over a range of detector trigger frequencies. The power consumption and ability of the firmware to successfully process this data was investigated. 

\section{MSP430FR5994 Firmware Operation}
\begin{figure*}[htbp]
\centerline{\includegraphics[width=0.99\linewidth]{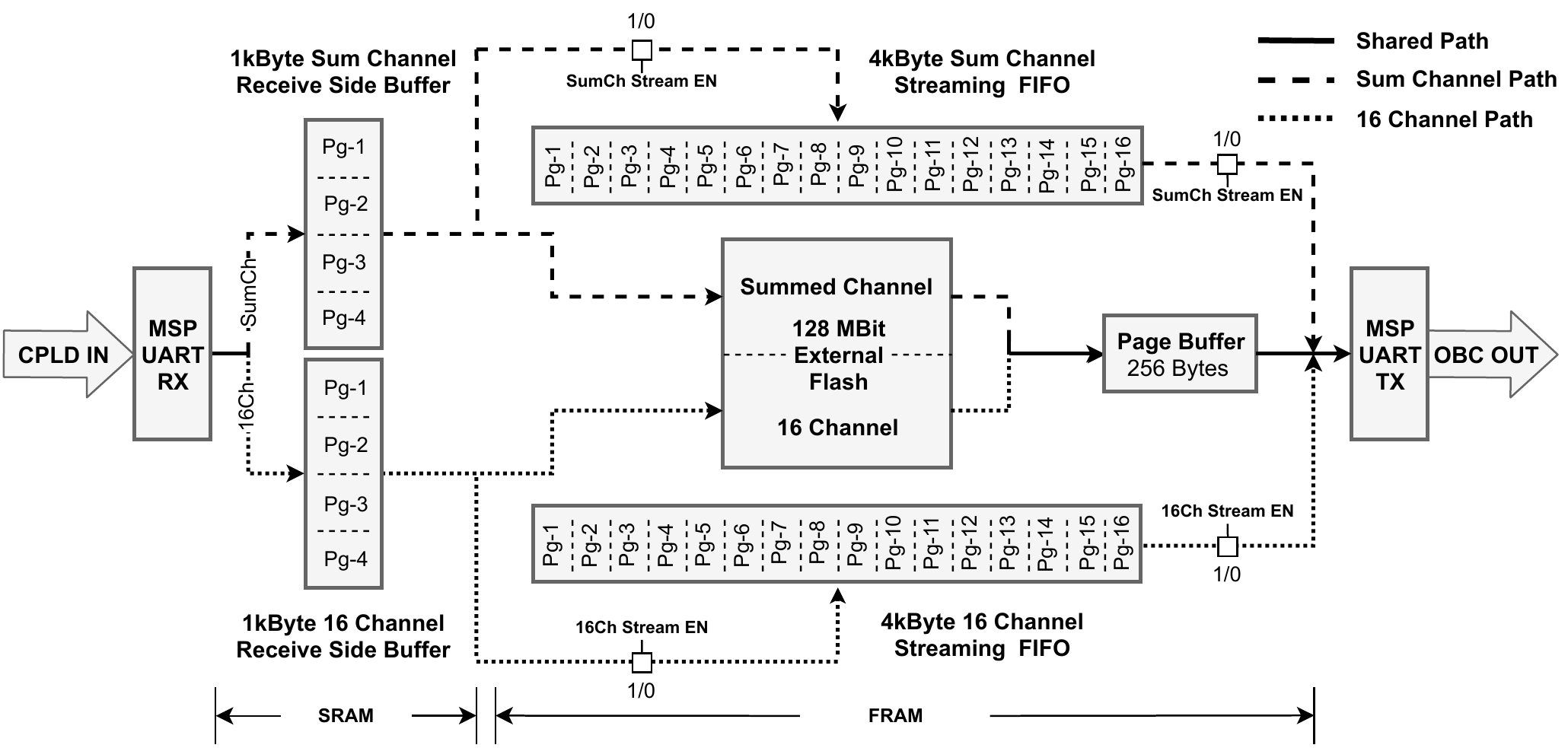}}
\caption{The \gls{gmod} experiment flow diagram, detailing reception of TTEs from the CPLD, processing and storage and finally service to the EIRSAT-1 on-board computer. TTE packets are received from the CPLD by the MSP430 and placed into the corresponding (summed or 16 channel TTE) receive side buffers. Once a full page of either buffer has been filled, its contents are passed to temporary storage in an external 128-Mbit flash until requested by the on-board computer. If streaming has been enabled, the filled page may also be placed within the corresponding streaming buffer for direct transmission to the on-board computer. The motivation behind the development of this initial version of the firmware was to provide an adequate platform to test the instrument and to investigate how best to ultimately manage the experiment operation using the MSP430. It is expected that this firmware will be updated and refined through further testing and development.}
\label{fig:GMODflow}
\squeezeup
\end{figure*}

In summary, the firmware operation is such that \gls{tte} packets are recieved from the \gls{cpld}, are processed and then stored in external 128-Mbit flash for service to the \gls{obc}. The MSP430 manages this flash storage by dividing it into a pair of circular ring buffers of equal size (a separate partition for summed channel and 16 channel \glspl{tte}). Upon request from the \gls{obc}, these \glspl{tte} are retrieved by the MSP430 and served to the \gls{obc} for \gls{grb} triggering, light curve and spectra generation. The MSP430 may also ``stream'' the \gls{tte} data directly to the \gls{obc} from internal memory for improved data transmission (all \gls{tte} data is stored in flash in the event that streaming becomes desynchronised, in this case servicing must restart from the last expected address). The MSP430 is also responsible for other duties which include the configuration of the \gls{asic}, \gls{cpld} reprogramming and enabling and controlling of the \gls{sipm} bias voltage and 80\,$\mu$A constant current supply (for internal \gls{asic} biasing). As a contingency while in flight, it is possible for the MSP430 to reprogram the \gls{cpld} using \gls{xsvf} instructions and a software interpreter which decodes the commands into the required \acrshort{jtag} I/O stimulus\cite{XAPP058_2017}. For brevity, only the experiment operation will be examined here as represented in Figure\,\ref{fig:GMODflow}.

The MSP430 receives \gls{tte} packets generated by the \gls{cpld} based on trigger data from the \gls{asic}. Interrupts are generated by the designated MSP430 serial module at the reception of individual bytes from the \gls{cpld}. The MSP430 waits until a recognisable 2 byte \gls{asm} sequence has been received which identifies the type of incoming \gls{tte} and will determine the specific response actions for summed or 16 channel packets. Both summed and 16 channel data are handled separately within the MSP430 memory and external flash and must be individually requested by the \gls{obc}. This is because summed channel \gls{tte} packets are the primary science product of \gls{gmod}; all scientific operations for \gls{grb} triggering will be carried out using summed channel data within the \gls{obc}. 16 channel \gls{tte} packets are a secondary product and have limited utility except for use in technology demonstrations or in health check assessments of individual \gls{sipm} channels. In flight, the main mode of operation will likely not include 16 channel readout, but it may be enabled on occasion.  

Once a recognised \gls{asm} has been received, the expected number of \gls{tte} serial bytes are read in and stored within a set of \gls{sram} circular buffers in MSP430 memory. The \glspl{tte} are temporarily stored here before transmission to flash or before being queued for streaming. There are two of these ``receive-side'' buffers on the MSP430, one for summed channel data and one for 16 channel data, both of identical length ($\sim$1\,kB). Each circular buffer can contain up to 4 pages (4$\times$256\,bytes, where 1 page = 256 bytes) of the given \gls{tte} data type. Each page can hold 31$\times$ summed channel \glspl{tte} or 7$\times$ 16 channel \glspl{tte}. This difference is due to the larger number of bytes per 16 channel \gls{tte} as compared to summed channel \glspl{tte}, effectively meaning that the 16 channel receive-side buffer is $\sim$4 times smaller in terms of the number of \glspl{tte} which can be packed within each 256 byte page. For this reason, at high data rates, the 16 channel buffers are expected to fill faster than the summed channel buffers. If at any point a receive-side buffers fills completely, \gls{tte} data is no longer accepted for the given \gls{tte} type, until at least a page worth of the received side buffer has been emptied by transmission to flash storage (this is a temporary solution employed only for initial testing).

Once a full page of either \gls{tte} type has been received, it may be stored within flash memory. The receive side buffer may continue to be filled during the writing of data to flash. A state within the main loop is entered which primes the page for transmission to flash storage; the flash command and address bytes are prepended to the contiguously arranged \gls{tte} page bytes,  a 4 byte coarse-timestamp is applied to the page with 1s time base, and for summed channel a \gls{crc} is calculated (due to the lower priority status of 16 channel data, error detection is not considered beneficial for this data type and is excluded). The page is then transmitted to flash over 8\,MHz SPI, using \gls{dma} to place each byte from \gls{sram} memory into the SPI module transmit buffer. \gls{dma} transfer of data has been used extensively in the experiment firmware, where possible, in an attempt to optimise the transmission of large amounts of \gls{tte} data and to reduce the chance of bottlenecking at high trigger rates. 

\gls{sram} memory has been used for the receive-side buffer storage; while \gls{fram} is more abundant on the MSP430FR5994, its read/write cycles are capped at 8\,MHz\footnote{It is possible to improve the write speed of \gls{fram}\cite{SLAA498B_2011} access in conjunction with \gls{dma} transfers, something which is exploited in other areas of the MSP430 firmware.} where as \gls{sram} is limited only by the clock source which in the current configuration is set to 16\,MHz. The use of \gls{sram} helps to ensure data in the receive-side is transferred to external flash with minimum overhead. As with the receive-side buffers on the MSP430, the Winbond W25Q128JV 128-Mbit flash is partitioned in half, the upper address space being reserved for summed channel data while the lower range being reserved for 16 channel data, as in Figure\,\ref{fig:address_scheme}. As the addressing for the flash is a 3 byte sequence, the boundaries between the two \gls{tte} data types can be easily identified by the firmware, using the most significant bit of the addressing to differentiate the summed channel range and the 16 channel range -- this is useful as a simple bit mask can be used to prevent accidental pointer overflow. Each flash partition acts as an independent circular buffer, similar to the receive-side buffers in the MSP430 \gls{sram} memory. In the case of the flash memory, a ``writing pointer'' for each buffer is recorded by the MSP430 which is an index of the address in flash to be written to next. This is as opposed to the conventional ``head'' and ``tail'' used in most circular buffers, which point to the start and end of the stored data respectively. The \gls{obc} can read this pointer and determine the amount of unread \gls{tte} pages in flash memory. It is important that the amount of unread data in flash which has yet to be transmitted to the \gls{obc} is kept as close to a minimum as possible to avoid the unlikely event that the flash buffer overflows and \gls{tte} data is overwritten. For the expected background rate of 50\,Hz in \gls{leo}, the summed channel buffer will take $\sim$5.7\,hours without reading to fill, with the 16 channel buffer filling within $\sim$1.2\,hours without reading. As the memory capacity available is so large, particularly in the summed channel buffer memory space (which is to be used for \gls{grb} triggering), it is unlikely that the \gls{obc} will need to request data from address outside all but a narrow window of address space.

As flash memory is used as a circular buffer, the memory address space is cycled many times during the experiment operation. Once the \gls{tte} data has been written to flash and is no longer needed in storage, rewriting to this location requires a pre-erase of the memory location essentially priming it for subsequent write operations. 
 
These erase operations take a considerable amount of time to complete (for example $\sim$45\,ms for 4\,kB sectors and $\sim$120\,ms for a 32\,kB block). When these erases are staged to occur in the MSP430 firmware, critical read/write operations to/from flash are blocked for this duration. Given these long erase times, it is possible that the receive-side buffers may overflow before the flash is available again for writing. Erases are therefore managed in a dynamic way by suspending flash erase operations to prioritise the writing of new \gls{tte} data to flash. Once the write has completed, the erase operation may then be resumed and continued in the background until completion. In the case of \gls{gmod}, erase operations are performed on a 32\,kB block basis, one block ahead of the ``write pointer" location ($\sim$128\,ms completion time). Thus the current block will have been cleared previously and be available for writing, while the following block has either been erased or is currently being erased, ensuring that there is always $\geq$32\,kB of flash available for writing and only a minimal amount of stored data is erased and recycled for reuse. In this way, the erase of an addressed block of flash memory is triggered by proximity to the current ``write pointer" address (ie. when writing to the start of a 32\,kB block boundary, the next block is immediately staged to be cleared by erasing its contents for reuse, as illustrated in Figure\,\ref{fig:address_scheme}). This strategy strikes a balance between maintaining as much historical TTE data in memory as possible while still ensuring that erases occur in a timely manner to allow new data to be successfully written.

\begin{figure}[t!]
\centerline{\includegraphics[width=0.84\linewidth]{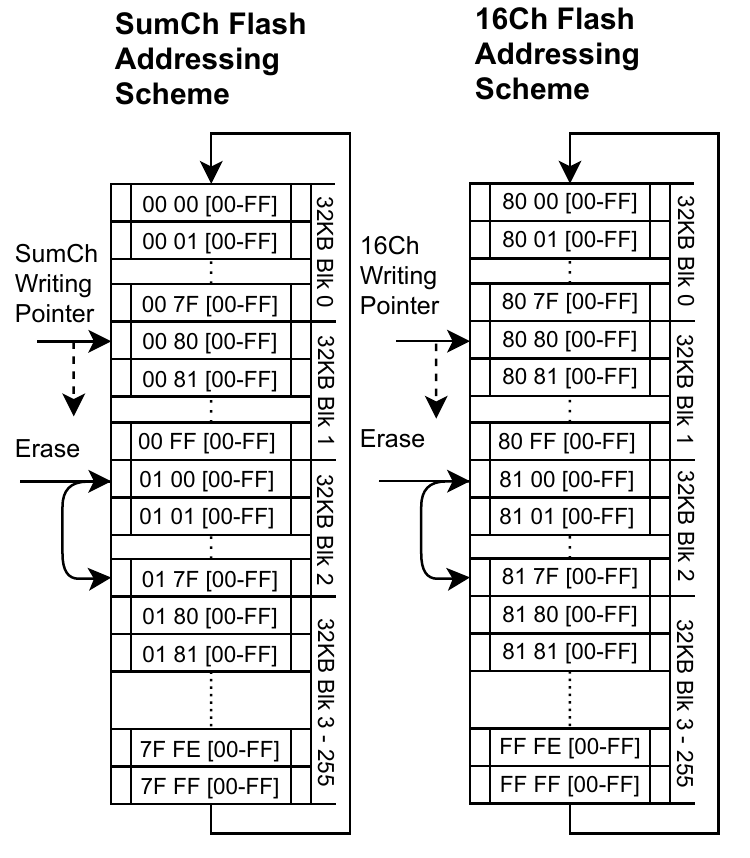}}
\caption{The external flash circular buffers used for summed and 16 channel \gls{tte} storage. This demonstrates the writing and erase procedure of flash  memory. Erasures occur when the writing pointer reaches the start of a previously erased 32\,kB block; the next 32\,kB block is erased in preparation for the subsequent writing of new TTE data. The \gls{obc} has unrestricted access to the flash and may request individual page reads from anywhere in memory or for streaming to begin from any address. The addressing scheme is also shown, where the first half of flash is reserved for summed channel data and the second for 16 channel data.}
\label{fig:address_scheme}
\squeezeup
\end{figure}

When the \gls{obc} requests data from a given address it is retrieved from flash in pages and placed within MSP430 \gls{fram} memory. This flash read operation is handled using \gls{dma} to read data from the \gls{eusci} SPI receive buffer and load it into \gls{fram} memory. A frame structure is applied to the page before sending to the \gls{obc}, with a header containing a unique 4 byte \gls{asm} sequence, identifying the start of the data frame and the flash address bytes from where the page originated. It is then transmitted to the \gls{obc} over 128\,kBaud serial. Once the \gls{obc} receives the packet and confirms the address bytes are as expected, the next page may then be requested if available.

For a faster throughput of data from \gls{gmod} to the \gls{obc}, a streaming system has been developed which allows the MSP430 to serve \gls{tte} data from internal memory rather than fetching from external flash. It is expected that this will be the primary mode of data transfer from the MSP430 during operation. When the MSP430 transfers the \gls{tte} page to flash (and when the streaming mode is enabled), that same page is transferred to a larger $\sim$4\,kB circular buffer which can hold 16$\times$ \gls{tte} pages. A pair of $\sim$4\,kB ``streaming'' buffers are reserved in \gls{fram}, one for summed channel data and another for 16 channel data. When a page has been loaded into either streaming buffer, it may be transmitted to the \gls{obc}. The \gls{obc} may then check the received address bytes in the packet frame to ensure that the correct page was sent by \gls{gmod} and that the data has been transmitted in the expected consecutive order according to the incriminating address bytes. If an inconsistency is detected between received \gls{tte} pages, the \gls{obc} may disable streaming and reenable streaming from the last expected page address in flash. The MSP430 then streams the contents directly from flash up from this address to the current ``writing pointer'' and the re-enables the streaming from internal memory at this point, whereupon subsequent pages are transmitted directly from the MSP430 \gls{fram}. If these streaming buffers overflow, the MSP430 can proactively reenable streaming from flash memory to the current ``writing pointer'' and then reenable streaming directly from the MSP430 \gls{fram}.       

Within the main loop of the firmware application is a set of states which define GMOD's operating configuration. At this stage of development, four distinct modes are defined, the two most relevant being \textit{Idle Mode}, where \gls{gmod} may be configured but where the experiment is not currently active and \textit{Experiment Mode}, similar to Idle Mode, except the experiment is currently running. The \gls{obc} may move \gls{gmod} between these different modes, which define certain configurations and the scope that the MSP430 can independently act within. To take full advantage of the low power optimisation on the MSP430, when \gls{gmod} is placed into Idle Mode, the MSP430 returns to a \gls{lpm}-3 state (where the \gls{cpu} is deactivated and the two high speed clock sources (out of three sources in total) have been disabled). In Experiment mode the MSP430 returns to a \gls{lpm}-0 state (where both the \gls{cpu} and its clock source are disabled). Deeper \gls{lpm} states allow low power operations at the expense of functionality, speed and device wake up time. For instance, placing the MSP430 into \gls{lpm}-0 while in Experiment Mode has the trade off of slightly higher power draw, while retaining the fast interrupt response and wake up time necessary for TTE collection.

\section{Initial Performance Analysis}

A number of iterations of this firmware have been developed and tested. Initial testing using the current firmware, indicates that the MSP430 is capable of meeting the requirements of the \gls{gmod} instrument in terms of readout and storage, power consumption and data rate handling in the context of \gls{grb} detection. The developed firmware must be able to successfully meet the following minimum operating requirements in order to be considered successful:

\begin{itemize}
    \item Retention of all received \glspl{tte} from SIPHRA and the \gls{cpld} without loss or corruption. 
    \item Managing trigger rates up to at least $\sim$500\,Hz for summed channel data (as this is used for \gls{grb} triggering).
    \item Achieve Low power optimisation with firmware development -- power must be less than 181.5\,mW (exc. 30$\%$ margin) estimated consumption as defined in the spacecraft initial \gls{ddf}\cite{ECSS_M_ST_10C_2009}. 
\end{itemize}

This firmware was tested by externally triggering SIPHRA at a range of fixed trigger rates, intercepting the \gls{cpld} output and comparing it against the MSP430 output. This was done in two separate tests, a test where only the summed channel readout was enabled and the other where both the summed and 16 channel readout were enabled. The intention behind this was to investigate the effects of bottlenecking within the MSP430 during high data rates when both 16 channel and summed channel \gls{tte} readout was enabled, to determine the trigger rate at which this would begin and to assess how many summed \glspl{tte} would be lost as a result. Furthermore, it was expected the power consumption of the MSP430 and \gls{gmod} motherboard would increase with both summed and 16 channel readout being processed, as at high data rates the MSP430 is likely being kept in a sustained active state while the flash is being read and written to more often than when just summed channel data is being processed. For this test only the reception of summed channel data was examined given its higher priority for scientific operations as opposed to 16 channel data. 16 channel data was also received to adequately simulate a full readout, however the handling of this data is beyond the scope of these initial tests and is not discussed.

An external hold input for SIPHRA's track and hold circuit is available, which when asserted, forces a readout which is processed by the \gls{cpld} and assigned a fine-timestamp as during normal operation. For this test, the hold signal was generated using the Aim-TTi TG5011 pulse generator, which was configured as a 3.3\,V, 25\,$\mu$s pulse width output, across a frequency range spanning 50\,Hz up to 1\,kHz. For the purposes of this test, the detector assembly, with the exception of the SIPHRA \gls{asic}, was not needed and was disconnected. A modified version of the \gls{cpld} \acrshort{vhdl} was used to mirror the serial output to the MSP430 over an easily accessible \acrshort{gpio} pin which could be logged to a PC using an FTDI USB to serial adaptor. \gls{gmod} was placed into Experiment Mode and commanded to begin streaming \gls{tte} data which was similarly also logged using a second \acrshort{ftdi} serial device. It was at this point that the readout configuration was set with streaming enabled (ie. summed channel readout or both summed and 16 channel readout). During the run, the current draw and voltage of the \gls{gmod} \gls{mb} 3.3\,V line, supplied by an Aim-TTi QL564T bench top \gls{psu} was monitored using a Rigol DM3058 digital multi-meter. The \gls{sipm} bias supply was disabled and was not used during this test. The current measurements were sampled 100 times for each run across the frequency range for summed channel only and summed and 16 channel readout cases. The \gls{tte} fine-timestamps were used to narrow the range of collected data to a burst of 30 seconds worth of \glspl{tte} at a given trigger rate. This was to attempt to simulate a burst from a long \gls{grb}, as the mean T$_{90}$ duration for long \glspl{grb}, according to the fourth Fermi \gls{gbm} catalogue is 29.9 seconds\cite{von_Kienlin_2020}. Each \gls{tte} from the \gls{cpld} data set was then searched for in the MSP430 received data to confirm that it was successfully processed by the MSP430.   

The results of this investigation are presented in Figures\,\ref{fig:percentage_dropped_ttes} and \ref{fig:power_draw}. Figure\,\ref{fig:percentage_dropped_ttes} shows the percentage number of lost \glspl{tte} for both summed channel readout and summed and 16 channel readout, both across the 50--1000\,Hz range, in incremented 25\,Hz steps below 200\,Hz and 100\,Hz increments above this range. For both readout configurations, it can seen that the firmware and MSP430 are capable of reliably receiving \gls{tte} packets from the \gls{cpld} and transmitting them to the \gls{obc} without losses up to $\sim$600\,Hz. This is well above the $\sim$120\,Hz trigger rate expected for a 10\,$\sigma$ significance trigger and the 500\,Hz requirement. This suggests that the firmware and MSP430 will be capable of reliably processing \gls{tte} data from the vast majority of \glspl{grb} without losses, even those few well above 10\,$\sigma$ significance. Above 600\,Hz, the percentage of lost \glspl{tte} begins to increase, showing a large difference between dropped \glspl{tte} for both readout configurations. As expected, when summed channel readout is selected with 16 channel readout, more \glspl{tte} are dropped as compared to the summed only configuration. This is because of the large quantity of data received especially for summed and 16 channel readout. It is expected that at high data rates, the receive-side buffers will overflow, at these times new \glspl{tte} will not be received from the \gls{cpld} and will be lost, prioritising already received data and its storage into external flash (as mentioned previously, this is a temporary solution for these initial tests). The MSP430 may also be in a state in which it cannot react to the reception of serial data from the \gls{cpld}, in these cases, it is possible for the \gls{asm} bytes to not be read and the \gls{tte} cannot be received. The MSP430 still shows good performance even at high trigger rates, particularly the summed channel configuration which is consistently below a fraction of a percent, even up to 1000\,Hz.

Figure\, \ref{fig:power_draw} shows the power consumption from the 3.3\,V line on the \gls{gmod} \gls{mb} for the same range of trigger frequencies. The left hand panel shows the power consumption when \gls{gmod} is in Idle mode (IDLE\_LPM3\text{*}) before the instrument configuration and just after programming, Idle mode (IDLE\_LPM3) after configuration and finally Experiment mode (EXP\_LPM0) before external triggering has begun. This highlights the use of \gls{lpm} states as being advantageous, particularly on CubeSat missions where good power management is a must. As expected, summed and 16 channel readout together consume more power, increasing as a function of trigger frequency. In both cases the power consumption is less than the original estimate of 181.5\,mW as defined in the \gls{ddf}. It is interesting to note that the curves both have a set of breaks around trigger frequencies which have been seen in past testing to be the points when the MSP430 streaming buffers overflow and servicing from flash begins. The summed channel streaming buffer was seen to overflow $\sim$600-700\,Hz -- a break can be seen in both curves around this range. For the summed and 16 channel configuration, a break is also seen around 200\,Hz -- previously the 16 channel streaming buffer was seen to overflow $\sim$200\,Hz. This is a somewhat unexpected, the reason why these notches appear is not yet known.  

\begin{figure}[t!]
\centerline{\includegraphics[width=0.95\linewidth]{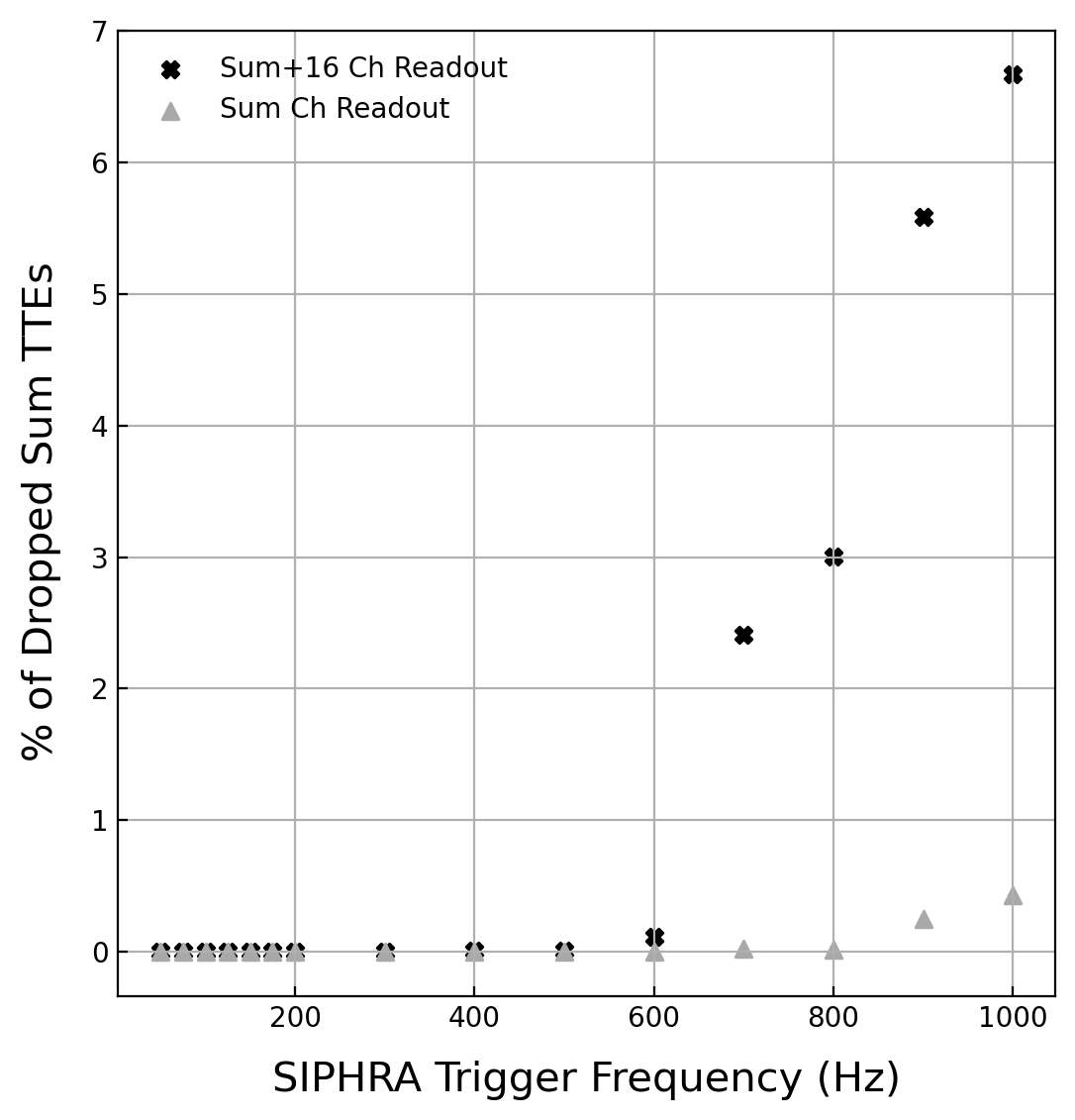}}
\caption{The percentage of dropped summed channel \glspl{tte} during summed and 16 channel readout and summed only readout, across a range of trigger rates.}
\label{fig:percentage_dropped_ttes}
\squeezeup
\end{figure}

\begin{figure}[t!]
\centerline{\includegraphics[width=1.05\linewidth]{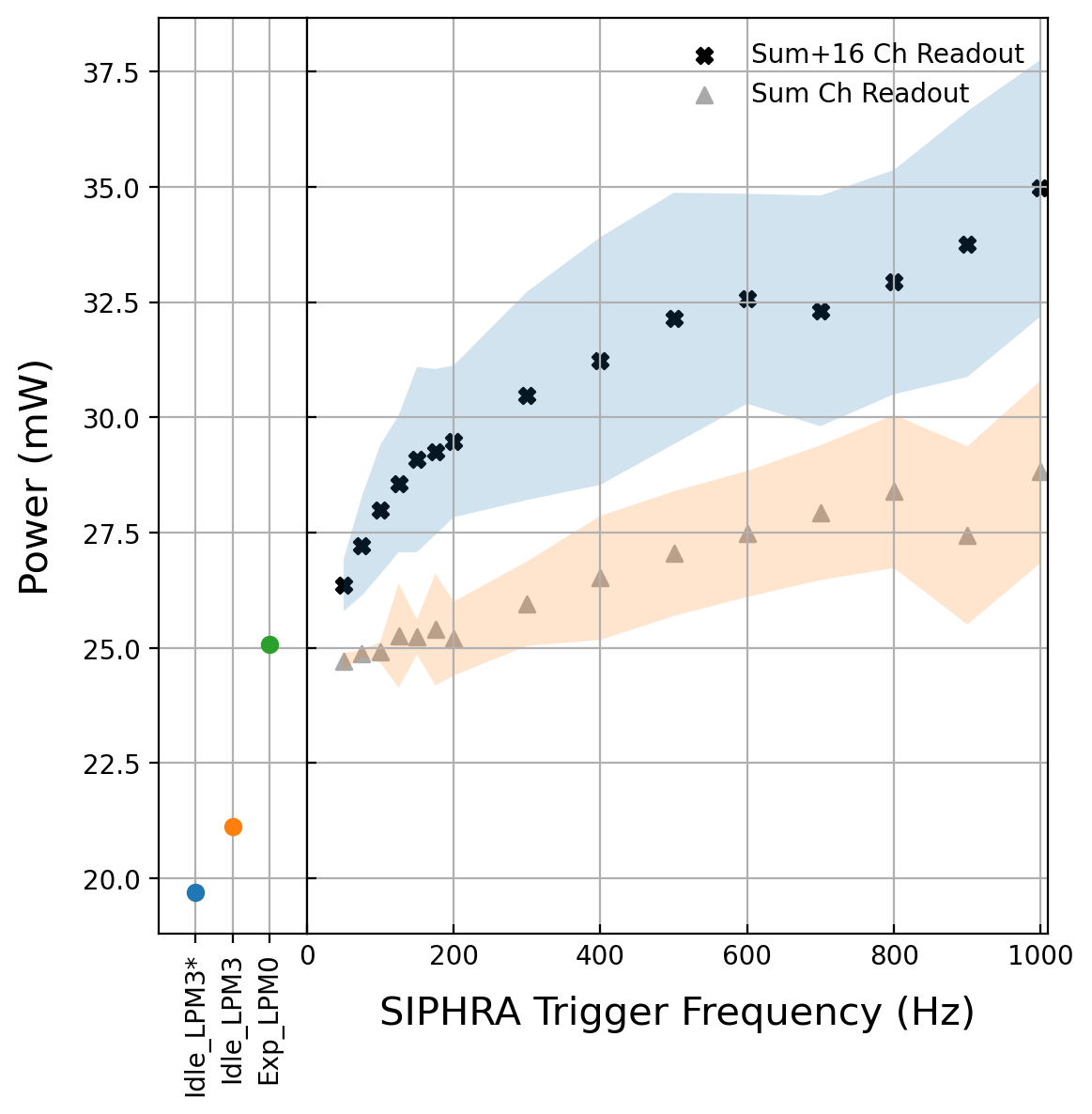}}
\caption{The power consumption derived from current measurements on the 3.3\,V line during testing. The fill region is the standard deviation calculated from 100 current measurements. The left hand panel shows the power consumption for Idle mode (IDLE\_LPM3\text{*}) before configuring, Idle mode (IDLE\_LPM3) after configuration and finally, Experiment mode (Exp\_LPM0) before external triggering.} 
\label{fig:power_draw}
\squeezeup
\end{figure}

There are a number of caveats and limitations to these initial tests. Firstly, as a fixed duration of 30 seconds of \gls{tte} data is being selected across a number of trigger rates, there will be a dissimilar number of \gls{tte} samples for each trigger rate data point. Secondly, the use of a pulse generator which triggers SIPHRA readout at a fixed period is not fully representative of a realistic scenario and does not account for \glspl{tte} which have been successively triggered by gamma-ray interactions. During testing, it has been noticed that with the current firmware, the MSP430 has a maximum ``dead time'' of around 15\,$\mu$s which is caused by receive-side pointer setup; after receiving a \gls{tte} from the \gls{cpld}, it is likely that any subsequent \gls{tte} whose \gls{asm} is received within this 15\,$\mu$s window will not be registered and will be lost. Thirdly, as the \gls{sipm} array and scintillator were not used for this test, the current measurements used to calculate the power consumption exclude any contributions from the bias generation circuit. It is also difficult to isolate the current draw of the MSP430 from the rest of the circuits of the \gls{mb} on the 3.3\,V line, circuits whose power consumption may be a function of trigger rate and \gls{tte} throughput. Finally, at the time of testing, the \gls{crc} and coarse-timestamp were not included in the firmware, an exclusion which may have an influence on \gls{tte} throughput to the external flash memory. 

\section{Progress and Next Steps}
The results presented here demonstrate the capability of the MSP430 to handle the transfer of data over the range required in orbit \cite{Murphy_2021_a} and that the power consumption is within the expected range for a variety of modes and experiment configurations. While the MSP430 performs to the required specifications for EIRSAT-1, with the future aim of scaling the \gls{gmod} instrument up, it is likely that a bespoke system-on-a-chip would be required to handle the rates expected from a larger instrument. Development is currently on going with this firmware, which is not yet considered flight ready, but will be further improved and developed with rigorous testing of specific loading cases and a verification in the full functional\cite{walsh_2020} and mission testing\cite{Doyle_2021} of the EIRSAT-1 Engineering Qualification Model and the Flight Model thereafter.


\section*{Acknowledgment}

The EIRSAT-1 project is carried out with the support of ESA’s Education Office under the Fly Your Satellite! 2 programme. Students acknowledge support from the Irish Research Council, the School of Physics and School of Computer Science in UCD.  We acknowledge all students who have contributed to EIRSAT-1. 
This study was supported Science Foundation Ireland (SFI) under grant number 17/CDA/4723 and by The European Space Agency's Science Programme under contract 4000104771/11/NL/CBi.  LH acknowledges support from SFI under grant 19/FFP/6777. DM, RD, and MD acknowledge support from the Irish Research Council (IRC) under grants GOIPG/2014/453, GOIPG/2019/2033 and GOIP/2018/2564 respectively. We acknowledge all students who have contributed to EIRSAT-1.

\bibliography{main}
\bibliographystyle{IEEEtran}
\end{document}